\documentclass[12pt]{iopart}
\usepackage{iopams}  

\usepackage{tikz-cd}
\usepackage{quiver}
\usepackage{float}
\usetikzlibrary{arrows}
\usepackage{amsfonts}
\usepackage{accents}
\usepackage{graphicx}				
\usepackage{amssymb}
\usepackage[all]{xy}

\usepackage{subfigure}
\usepackage{subcaption}
\usepackage{varwidth}

\expandafter\let\csname equation*\endcsname\relax
\expandafter\let\csname endequation*\endcsname\relax

\usepackage{amsmath}

\begin{document}

\title[Physical computation and system compositionality]{Physical Computing:\\ A Category Theoretic Perspective on Physical Computation and System Compositionality}

\author{Nima Dehghani$^1$,$^2$ \footnote{Present address: McGovern Institute for Brain Research, MIT, Cambridge, 02139, MA, USA} 
and Gianluca Caterina$^3$,$^4$}

\address{$^1$ Department of Physics, MIT, Cambridge, 02139, MA}
\address{$^2$ McGovern Institute for Brain Research, MIT, Cambridge, 02139, MA, USA}
\address{$^3$ Department of Mathematics, Endicott College, Beverly, MA, 01915, USA}
\address{$^4$ Center of Diagrammatic and Computational Philosophy, Endicott College, Beverly, MA, 01915, USA}
\ead{nima.dehghani@mit.edu}


\vspace{10pt}
\begin{indented}
\item[]Dec 2023
\end{indented}

\begin{abstract}
This paper introduces a category theory-based framework to redefine physical computing in light of advancements in quantum computing and non-standard computing systems. By integrating classical definitions within this broader perspective, the paper rigorously recontextualizes what constitutes physical computing devices and processes. It demonstrates how the compositional nature and relational structures of physical computing systems can be coherently formalized using category theory. This approach not only encapsulates recent formalisms in physical computing but also offers a structured method to explore the dynamic interactions within these systems.
\end{abstract}

%
\vspace{2pc}

%
%
%
\section{Introduction}
Roots of computability trace back to Leibniz, who invented a mechanical calculator for automating the evaluation of mathematical expressions \cite{Bell1986,Davis2018,Chaitin2007}. At the Paris international conference for mathematics, David Hilbert extended Leibniz's fascination by proposing the \emph{Entscheidungsproblem} (the decision problem), questioning the existence of an ``effective procedure'' \footnote{Prior to the emergence of algorithm(s), procedural calculation through a finite number of exact, finite instructions was labeled ``effective procedure (or effective calculation).} to determine the truth or falsity of mathematical statements \cite{Hilbert1902}.

Alan Turing and Alonso Church independently demonstrated the impossibility of resolving the \emph{Entscheidungsproblem}. This discovery, known as the ``Church--Turing thesis'', posited that no effective procedure (or ``algorithm'' in contemporary terms) can perform these calculations. Their approach involved first ``precisely''  defining ``effective procedure (effective calculation)'', thus formalizing ``computable function''. Next, they established that no computable function can solve Hilbert's problem. In his contribution, Church introduced $\lambda$-calculus and proved the inability to determine equivalence between $\lambda$-calculus expressions \cite{Church1936}. Turing developed the concept of ``Turing machines'', showing the impossibility of a universal method to decide the halting of any given Turing machine \cite{Turing1937}.

The exploration of these ideas led to the inception of Turing machines, computability theory, digital computers, and subsequent advances in the information industry. The foundational mathematical nature of computability theory ensures that various computational models like ``state model'' (e.g., Turing Machine  or Finite State Machines), ``functional'' (Lambda calculus), ``logical'' (Logical programming), or ``concurrency'' (e.g., Process calculus or Petri nets) are effectively equivalent \cite{Cooper2017}. The directional arrow from computable functions to models of computation, to devices that can carry such models of computation, is precise and concrete. However, applying this trajectory in reverse from {\em Natural systems} to {\em operations} to {\em mathematical abstraction} introduces ambiguity and imprecision. This paper addresses: \emph{What physical systems can be considered computers?} and \emph{What constraints enable a physical system to compute?}, building upon the insights of \cite{Horsman2014, Horsman2018} and introducing a category theory-based formalism.

\section{What is a Physical Computer?}
Reversing the trajectory from digital computing to physical systems unearths various philosophical and practical questions in computational theory. One line of inquiry questions whether simply a change in the state of physical objects can be deemed as computing, such as a rock calculating its trajectory \cite{Chalmers1996}. Similarly, this perspective extends reducing the interaction between physical objects to the notion of computation, prompting discussions such as whether the solar system computes natural laws \cite{Campbell2021}. In biological contexts, where the governing laws are less defined, these questions gain vague intricacies. Discussions in this realm range from general acceptance that biological systems perform computations, to more specific inquiries about the nature and types of biological computations and algorithms present in simpler organisms like bacteria or complex ones like the human brain; posing as \emph{what is (biological) computation?} \cite{Mitchell2011} or asking \emph{what type of biological algorithms are out there?} \cite{Schnitzer2002}, \emph{do bacteria compute?} \cite{Benjacob2009} or \emph{is brain a computer?} \cite{Richards2022,Brette2022}.

The challenge arises since the answer to these types of questions are shaped based on the trajectory from computable function to computational model to computer 
while our (incomplete and partial) knowledge about these systems is limited to their physical construct and interaction. A theoretical definition of computation exists \cite{Copeland1996}, yet the application to natural systems requires a nuanced approach. To avoid simplistic analogies and to explore the true computational nature of various systems, it's necessary to develop a formalism that can effectively navigate from natural systems to operations and then to mathematical abstraction.  In order not to fall into the trap of misguided and perhaps sometimes useful vague analogies like ``X is a computer because it handles information'' or ``X is a computer because it processes information'' or ``X is a computer because it has state transitions'', we need to develop a formalism that can help us traverse the reverse trajectory, i.e. {\em Natural systems} $\rightarrow$ {\em operations} $\rightarrow$ {\em mathematical abstraction}.

\subsection{Physical Computation: A Representational Perspective}
Horsman et al.'s Abstraction/Representation (AR) theory, inspired by quantum computation, addresses a fundamental question in computational theory: when do physical systems compute? They propose a concept called the ``compute cycle'' \cite{Horsman2014,Horsman2018}. In this theory, physical computing is conceptualized as using a physical system to predict the evolution of an abstract representation. Central to AR theory is the utilization of commuting diagrams. These diagrams represent a bi-directional mapping between a physical system and its abstract counterpart. For instance, a physical system $p$ is abstractly represented as $m_p$ through the modelling representation relation $R_T$ of theory $T$. The abstract dynamics $C_T$ evolve the abstract state $m_{p}$ to $m_{p'}$, paralleled by the evolution of the physical system $p$ to state $p'$ by physical dynamics $H(p)$. This framework offers a method to visualize and understand the interaction between physical and abstract computational processes (as shown in Fig.\ref{fig:Horsman}).

\begin{figure}[hb]
  \centering
   \begin{subfigure}
     \centering
     \begin{varwidth}{\linewidth}
\[\begin{tikzcd}
	{m_p} &&& {m'_p} \\
	&&& {m_{p'}} \\
	\\
	p &&& {p'}
	\arrow["{C_T}", tail, from=1-1, to=1-4]
	\arrow["{R_T}", tail, from=4-1, to=1-1]
	\arrow["{H(p)}"', tail, from=4-1, to=4-4]
	\arrow["{R_T}"', tail, from=4-4, to=2-4]
	\arrow[color={rgb,255:red,214;green,92;blue,92}, dotted, tail, two heads, from=2-4, to=1-4]
\end{tikzcd}\]
\end{varwidth}
     \caption*{Theory-Experiment}
     \label{fig:diagram}
   \end{subfigure}
   \hfill
   \begin{subfigure}
     \centering
\[\begin{tikzcd}
	{m_p} && {m_{p'}} \\
	p && {p'}
	\arrow[tail,from=2-1, to=1-1,"R_T"]
	\arrow[dashed,tail,from=2-1, to=2-3,"H(p)",color={rgb,255:red,92;green,92;blue,214}]
	\arrow[tail,from=1-3, to=2-3,"\widetilde{R_T}"]
	\arrow[tail, from=1-1, to=1-3,"C_T"]
\end{tikzcd}\] 
     \caption*{Predict cycle}
     \label{fig:theory}
   \end{subfigure}
   \hfill
   \begin{subfigure}
     \centering
\[\begin{tikzcd}
	{m_p} && {m_{p'}} \\
	p && {p'}
	\arrow[tail,from=1-1, to=2-1,"\widetilde{R_T}"]
	\arrow[tail,from=2-1, to=2-3,"H(p)"]
	\arrow[tail,tail,from=2-3, to=1-3,"R_T"]
	\arrow[dashed,from=1-1, to=1-3,"C_T",color={rgb,255:red,92;green,92;blue,214}]
\end{tikzcd}\] 
     \caption*{Compute cycle}
     \label{fig:compute}
   \end{subfigure}
    \caption{Commuting diagram and representation of physical-abstract as proposed by \cite{Horsman2014,Horsman2018}. (a) Commuting diagram for an experiment to test a theory. (b) Abstract theory is used to predict the evolution of the physical system. (c) Evolution of the physical system (computer) is used to predict abstract evolution. Note: dashed blue lines in b and c are not utilized in predict and compute cycles.}
    \label{fig:Horsman}
\end{figure}

AR theory posits that a physical system is computing when the evolution of its abstract representation corresponds with 
abstract representation of the evolved physical system itself. This theory, articulated by Horsman et al., relies on representational mapping between physical and abstract domains, symbolized through $R_T$ and $\tilde{R_T}$, and illustrated using commutative diagrams. However, AR risks falling into the same conceptual pitfalls as Putnam's ``mapping account,''  \footnote{Later, referred to as the ``simple mapping account'' by \cite{Godfreysmith2009}. The simple mapping account requires two conditions in order for a physical system P to compute the descriptions in C \cite{Piccinini2010}:
\begin{enumerate}
  \item A mapping from the computational description (C) to a physical description (state P) 
  \item states and transitions of the physical system (P) should mirror the computational (C) states and transitions 
\end{enumerate}} which suggests that computation occurs in any physical system with a one-to-one mapping to a computational process \cite{Putnam1960,Putman1964,Putnam1988}. All what is needed to satisfy this condition is for the micro-states of the system to match the transitions specified by machine table (equivalent of the abstract computation) on hand. This perspective could lead to an overly broad definition of computational systems, lacking specific criteria to discern true computational processes.
An schematic of the simple mapping account is represented in \ref{fig:Simple} showing the correspondence between the abstract and the physical domains.

\begin{figure}[hb]
     \centering
\[\begin{tikzcd}
	& {m_p} &&& {m_p'} \\
	{} &&&&& {} \\
	& p &&& {p'}
	\arrow["Abstract"', tail, from=1-2, to=1-5]
	\arrow["Physical", tail, from=3-2, to=3-5]
	\arrow[color={rgb,255:red,153;green,92;blue,214}, dashed, no head, from=2-1, to=2-6]
\end{tikzcd}\]
        \caption{Simple mapping account (after Putnam) }
        \label{fig:Simple}
\end{figure}

The ``simple mapping account" posited by Putnam is challenged by its generality, as it implies that any physical system could be considered a Turing machine due to the vast number of its possible micro-states. This leads to a form of pancomputationalism, where virtually all physical systems could be seen as implementing some form of computation \cite{Piccinini2021}\footnote{Unpublished, personal communication \cite{Anderson2023}.}. Such a ``weak mapping" does not provide stringent criteria for physical-computational equivalence, opening the door to overly inclusive definitions of computation. Dennett's suggestion of reverse engineering mapping as a means to discern genuine computational patterns aims to move beyond the overly broad interpretations of computational systems \cite{Joslin2006}, but the key challenge remains in the overly general mapping of physical states to computational processes.

Mapping accounts, critical in understanding computation, have evolved with varying philosophical emphases. Versions like simple \cite{Putman1964}, causal \cite{Chalmers1994}, counterfactual \cite{Scheutz1999}, dispositional \cite{Klein2008}, syntactic \cite{Fodor1975}, semantic \cite{Shagrir2006}, and mechanistic \cite{Piccinini2007} were all developed to interpret the computational aspect of the human mind rather than any general computing physical system. In contrast, AR theory \cite{Horsman2014} was devised for unconventional computing, differing from traditional accounts by not confining abstract representations to mental aspects \footnote{For an in-depth review of different mapping accounts, see \cite{Piccinini2010}. For a review on relations between AR and different mapping accounts, see \cite{Fletcher2018}.}.  Weak forms of the mapping account (as referred to by \cite{Piccinini2021}) only require an arbitrary physico-computational association for the first condition to hold. That the association between a given physical system and a computation can be entirely based on mapping, opens the gate to pancomputationalism \cite{Piccinini2021}. AR theory relies on this physico-computational mapping; however, it binds the mapping to a representational frame. As a result, according to AR, there is no computation without such representation. This ``representational'' view of concrete computation is contrary to the simple \cite{Putnam1960} and mechanistic \cite{Piccinini2007} mapping accounts, and more in-line with the syntactic \cite{Fodor1975} and semantic \cite{Shagrir2006}. What sets apart AR from syntactic \cite{Fodor1975} and semantic \cite{Shagrir2006} accounts is that AR deviates from requiring the representational entity to be an aspect or product of mind or for the representations to be meaningful \cite{Fletcher2018}. In fact, AR does not put any restrictions on the abstract (representational) mapping of the physical state \cite{Horsman2014}. AR theory's unique approach sets it apart from ``philosophical'' account of concrete computation, but also presents challenges in defining computational activity, which we aim to resolve through a category-theoretic functorial perspective.

\section{Physical  computation: a functorial perspective}
As discussed above, various mapping account of the notions of concrete computation grapple with issues that either stem from pre-loaded notions of mind or lead to pancomputationalism. Notably, the mechanistic mapping account \cite{Piccinini2007,Piccinini2010} and its recent update ``the robust'' mechanistic account \cite{Piccinini2021} attempt to overcome these issues by resorting to a teleological description of concrete computation. The abstract representation of physical systems (AR) framework \cite{Horsman2014,Horsman2018}, while attempts to define a physical computational system from a non-subjective perspective, resorts to vague descriptions of representational mapping, therefore inheriting a mix of issues of the mapping accounts. According to AR, a physical system is deemed a computer if it meets several criteria:
\begin{enumerate}
  \item A theory $\mathcal{T}$ of the physical computational device. \label{var:a1}
  \item A bidirectional representational mapping between the physical system and the abstract domain. $\widetilde{\mathcal{R_T}}(a)$  representing the initial state of the physical system (which is encoded by the abstraction $a$). $\mathcal{R_T}(p')$ revert the final state of the physical system back to the abstract domain (they refer to this as decoding). 
  \item A set of one or more fundamental physical operations that transform input states to output states. 
  \item The components  i.e. theory, representation and operations must commute in a specific sequence (as shown in Fig \ref{fig:Horsman}).
\end{enumerate}

The AR framework's definition prompts several questions: a) is it compositional? b) what's the relationship between different abstract representations of the same system? c) can the inverse relationship of ``vertical maps" ($\mathcal{R_T}$ and $\widetilde{\mathcal{R_T}}$) be generalized? d) does it imply a causal relationship between the physical and abstract? e) can AR represent counterfactual dynamics (transitions between states)? Addressing these requires a precise mathematical framework, as current understanding of AR's compositionality and its representation relations lacks clarity. To tackle these issues, we propose a category theory-based framework, offering a rigorous structure to explore the interactions between physical and mathematical computing levels, and to define necessary conditions for physical computation.




\subsection{Categorical definitions}
Referencing Coeke and Paquette's work \cite{Coecke2011}, we define the category {\bf PhysProc} for concrete physical systems:

\begin{itemize}
\item Physical systems $A, B, C, \dots$ are considered as objects.
\item Morphisms are processes transforming a system of type A into type B, denoted as $A\rightarrow B$.
\item These processes typically need finite time to complete.
\item Sequential composition of processes is treated as composition, with identity processes leaving the system invariant.
\end{itemize}

Demonstrating {\bf PhysProc} as a \emph{category} is straightforward (detailed in the Appendix).

In \cite{Coecke2011}, various real-world, concrete, and abstract categories are defined, focusing on quantum processes. Our approach, however, is broader, aiming to define an abstract category of representation that encapsulates the computational nature described in Horsman's work, without being restricted to specific physical processes.




We are introducing a category, denoted as \textbf{Comp}, to encapsulate abstract representations. This category, \textbf{Comp}, is pragmatically defined with the following components:

\begin{itemize}
\item \textbf{Objects}: All terms of data types, such as Booleans, integers, reals, etc., are considered as objects.
\item \textbf{Morphisms}: All programs which take data of term $A$ as their input and produce data of term $B$ as their output are considered as morphisms, denoted as $A\rightarrow B$.
\item \textbf{Composition}: The sequential composition of programs and the programs which output their input unaltered are considered as composition.
\end{itemize}

Additionally, we incorporate a \textbf{null} object into \textbf{Comp}, denoted as $0_{Comp}$. This null object is both an initial and a final one, serving as a placeholder for those physical systems that do not have an abstract representation under a given abstraction $\mathcal{R}_T$. The choice of \textbf{AbsProc} depends on the level of generality we want to represent computation, and \textbf{Comp} is a pragmatic and natural choice for this purpose.




We define {\em physical computation} as a pair of functors ($\mathcal{R_T}, \widetilde{\mathcal{R}}_T$) linking {\bf PhysProc} and {\bf AbsProc}. 

{\textbf {Definition:}} {\em A physical computation}   is a pair of functors ($\mathcal{R_T}, \widetilde{\mathcal{R}}_T$)

\[\mathcal{R_T}: {\bf PhysProc} \longrightarrow {\bf AbsProc}\]
\[\widetilde{\mathcal{R}_T}: {\bf AbsProc} \longrightarrow {\bf PhysProc}\]
These functors satisfy the conditions: 
\[\mathcal{R_T}\circ \widetilde{\mathcal{R}}_T=Id_{{\bf AbsProc}}\]
and 
\[\widetilde{\mathcal{R}_T}\circ \mathcal{R}_T=Id_{{\bf PhysProc}}\]
, where $Id_C$ is the identity functor on category $C$. This precise definition within a category theory framework aims to clarify the concept of computational physical systems.




This definition connects the `amount' of computation a physical system can perform to the `size' of the image of the functor $\mathcal{R_T}$. Systems capable of universal computation would result in a maximal $\mathcal{R_T}$, while systems not allowing any computation would lack such a functor.

In this context, the previously discussed commuting diagram is formalized, depicting the evolution process $f$ from a physical system $p$ to $p'$ and its corresponding computation in the abstract domain.
\[\begin{tikzcd}
	{\mathcal{R}(p)} && {\mathcal{R}(p')} \\
	p && {p'}
	\arrow[tail,from=2-1, to=1-1,"\mathcal{R}_T"]
	\arrow[tail,from=2-1, to=2-3,"f"]
	\arrow[tail,tail,from=2-3, to=1-3,"\mathcal{R}_T"]
	\arrow[from=1-1, to=1-3,"\mathcal{R}_T(f)",color={rgb,255:red,92;green,92;blue,214}]
\end{tikzcd}\] 
To complete the computing cycle, ensuring physical systems serve as viable computers, $\mathcal{R}_T$ should ideally have $\widetilde{\mathcal{R}}_T$ as its inverse, aligning the physical and abstract computational processes. This would allow an observer to use the physical system as a computer.





\[\begin{tikzcd}
	{\mathcal{R}(p)} && {\mathcal{R}(p')} \\
	p && {p'}
	\arrow[tail,from=1-1, to=2-1,"\widetilde{\mathcal{R}}_T"]
	\arrow[tail,from=2-1, to=2-3,"f"]
	\arrow[tail,tail,from=2-3, to=1-3,"\mathcal{R}_T"]
	\arrow[from=1-1, to=1-3,"\mathcal{R}_T(f)",color={rgb,255:red,92;green,92;blue,214}]
\end{tikzcd}\] 
Note that our approach is not merely a renaming of entities in the definition of computing devices but a mathematical formalization of computing cycles within suitable categories. This formalization brings several benefits:

\begin{itemize}
\item \textbf{Compositionality}: The categorical framework naturally captures the structured and compositional nature of physical computing processes. This emphasizes the inherent structure in these processes.
\item \textbf{Relations between different computations and theories}: The relations between different abstract representations of the same physical object, as well as the relations between different theories, are elegantly expressed through natural transformations (see Appendix for the notion of \emph{Natural Transformation}). This is crucial for understanding consistent computational representations across various theories. In other words, if two different representations, say $\mathcal{R}_T$ and $\mathcal{R}_T'$ , consistently perform the same computation, they should have some fundamental structural properties in common. This condition is satisfied in our framework by requiring the existence of a natural transformation, which is a morphism in the functor category of all the functors from \textbf{PhysProc} to \textbf{AbsProc}.
\item \textbf{Computational Refinement}: The concept of computational refinement is reinterpreted as a special case of natural transformations. This provides a new perspective on the process of refining computations.
\item \textbf{Flexible Representation of Computational Systems}: The requirement for $\mathcal{R}_T$ to have an inverse can be relaxed. Instead, we can require that the pair $(\mathcal{R}_T, \widetilde{\mathcal{R}}_T)$ forms an adjoint pair (see Appendix for the notion of \emph{Adjoint Pair}). This allows for a more flexible, `locally’ identity-based approach, thereby expanding the potential scope of what can be considered a computational system. 
\end{itemize}

\subsection{Compositionality in computation category}
\subsubsection{Categorical Composition:}
In the category framework, the compositional nature of morphisms is fundamental. For arrows between objects $A$ and $B$, and $B$ and $C$, there exists a \emph{unique} arrow $g\circ f$ from $A$ to $C$.


Formally, given two arrows between objects $A$ and $B$
\[\xymatrix{
A\ar[rr]^{f} & & B
}
\]
\[\xymatrix{
B\ar[rr]^{g} & & C
}
\]

there exists one and only one arrow
\[\xymatrix{
A\ar[rr]^{g\circ f} & & C
}
\]

This arrow $g\circ f$ is called the {\it composite} of $f$ and $g$. 

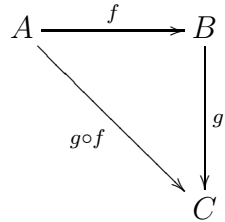
\[\xymatrix{
A\ar[rr]^{f}\ar[ddrr]_{g\circ f} & & B\ar[dd]^{g}\\
& & \\
& & C
}
\]

This composite arrow concept is intuitively represented through categorical diagrams, highlighting the sequential nature of computation in physical processes. Specifically, in {\bf PhysProc}, objects $p, p', p''$ and their evolution processes can be composed, reflecting the computational cycles in this framework.

That is, given three objects $p, p', p''$ in ${\bf PhysProc}$ such that 
\[p\rightarrow p'\rightarrow p''\]
along with the corresponding computing cycles

\[\begin{tikzcd}
	{\mathcal{R}(p)} && {\mathcal{R}(p')} \\
	p && {p'}
	\arrow[tail,from=2-1, to=1-1,"\mathcal{R}_T"]
	\arrow[tail,from=2-1, to=2-3,"f"]
	\arrow[tail,tail,from=2-3, to=1-3,"\mathcal{R}_T"]
	\arrow[from=1-1, to=1-3,"\mathcal{R}_T(f)",color={rgb,255:red,92;green,92;blue,214}]
\end{tikzcd}\] 

and

\[\begin{tikzcd}
	{\mathcal{R}(p')} && {\mathcal{R}(p'')} \\
	p' && {p''}
	\arrow[tail,from=2-1, to=1-1,"\mathcal{R}_T"]
	\arrow[tail,from=2-1, to=2-3,"g"]
	\arrow[tail,tail,from=2-3, to=1-3,"\mathcal{R}_T"]
	\arrow[from=1-1, to=1-3,"\mathcal{R}_T(g)",color={rgb,255:red,92;green,92;blue,214}]
\end{tikzcd}\] 

we have the composition

\[\begin{tikzcd}
	{\mathcal{R}(p)} && {\mathcal{R}(p'')} \\
	p && {p''}
	\arrow[tail,from=2-1, to=1-1,"\mathcal{R}_T"]
	\arrow[tail,from=2-1, to=2-3,"g\circ f"]
	\arrow[tail,tail,from=2-3, to=1-3,"\mathcal{R}_T"]
	\arrow[from=1-1, to=1-3,"\mathcal{R}_T(g\circ f)",color={rgb,255:red,92;green,92;blue,214}]
\end{tikzcd}\] 
To enhance clarity, we focus on the functor $\mathcal{R}_T$ in our diagrams while acknowledging the role of the inverse functor $\widetilde{\mathcal{R}}$. This addresses the need for a comprehensive representation of computational processes within the category theory framework.

Our approach ensures that for all $p\in{\bf PhysProc}$ and $a\in{\bf AbsProc}$, the compositions $(\widetilde{\mathcal{R}}\circ\mathcal{R})(p)=p$ and $(\mathcal{R}\circ\widetilde{\mathcal{R}})(a)=a$ hold, further solidifying the computational structure we propose.

\subsubsection{Refinement $\&$ Hierarchical Composition Through Natural Transformation:}

Exploring categories at a higher level, we consider those categories whose objects are functors and whose morphisms are `relations of relations', known as natural transformations (see Appendix). This is crucial for abstract representations performing similar computations, implying structural similarities.

For functors $\mathcal{R}T$ and $\mathcal{R}{T'}$, a natural transformation between them implies that for any physical systems $p, p'$, there exist abstract dynamics $\eta(p)$ and $\eta(p')$ forming a commuting diagram:

\[
\xymatrix{
\mathcal{R}_T(p)\ar[rr]^{\mathcal{R}_T(f)}\ar[dd]_{\eta_{p}} & & \mathcal{R}_T(p')\ar[dd]^{\eta_{p'}}\\
 & & \\
\mathcal{R}_{T'}(p)\ar[rr]^{\mathcal{R}_{T'}(f)} & &  \mathcal{R}_{T'}(p') 
}
\]

For example, the relation between decimal, octal, and binary adders can be represented through natural transformations, demonstrating the flexibility and depth of this categorical approach.




The concept of compositionality extends the functorial perspective from {\bf PhysProc} to {\bf AbsProc}, allowing for the creation of complex representations. This can be seen as a hierarchical composition in {\bf AbsProc} that broadens the functorial mapping between {\bf PhysProc} and {\bf AbsProc} to various abstraction levels, embodying the idea of ``refinement".

\begin{figure}[H]\scalebox{.5}
     \centering
\begin{tikzcd}
	{} &&&& {} &&&& {} \\
	&&& {6,9} & {} & 15 \\
	& {} &&&&&& {} \\
	&&& {{ 6 :\,11}} & { } & 17 \\
	&& {} &&&& {} \\
	&&& {110,1001} & {} & 1111 \\
	&& {} && {} && {} \\
	&&& {p_1} && {p_2} \\
	&& {} &&&& {} \\
	& {} &&&&&& {} \\
	{} &&&&&&&& {}
	\arrow["{{\:Binary}}"', from=6-4, to=6-6]
	\arrow[color={rgb,255:red,92;green,92;blue,214}, curve={height=-12pt}, no head, from=3-2, to=3-8]
	\arrow["{{\:Decimal}}", from=2-4, to=2-6]
	\arrow[color={rgb,255:red,214;green,92;blue,92}, curve={height=-12pt}, no head, from=1-1, to=1-9]
	\arrow[from=2-4, to=4-4]
	\arrow[from=4-4, to=6-4]
	\arrow[from=6-6, to=4-6]
	\arrow[from=4-6, to=2-6]
	\arrow[color={rgb,255:red,92;green,214;blue,92}, curve={height=-12pt}, no head, from=5-3, to=5-7]
	\arrow[from=8-4, to=8-6]
	\arrow[from=6-4, to=8-4]
	\arrow[from=8-6, to=6-6]
	\arrow[color={rgb,255:red,92;green,214;blue,92}, curve={height=-12pt}, no head, from=9-3, to=5-3]
	\arrow[color={rgb,255:red,92;green,214;blue,92}, curve={height=12pt}, no head, from=9-7, to=5-7]
	\arrow[color={rgb,255:red,92;green,92;blue,214}, curve={height=-12pt}, no head, from=10-2, to=3-2]
	\arrow[color={rgb,255:red,92;green,92;blue,214}, curve={height=12pt}, no head, from=10-8, to=3-8]
	\arrow[color={rgb,255:red,214;green,92;blue,92}, curve={height=12pt}, no head, from=11-9, to=1-9]
	\arrow[color={rgb,255:red,214;green,92;blue,92}, curve={height=-18pt}, no head, from=11-1, to=1-1]
	\arrow["{Decimal\:Computer}"{description}, color={rgb,255:red,214;green,92;blue,92}, curve={height=12pt}, no head, from=11-1, to=11-9]
	\arrow["{Binary\:Computer}"{description}, color={rgb,255:red,92;green,214;blue,92}, curve={height=12pt}, no head, from=9-3, to=9-7]
	\arrow["{Octal\:Computer}"{description}, color={rgb,255:red,92;green,92;blue,214}, curve={height=12pt}, no head, from=10-2, to=10-8]
	\arrow[color={rgb,255:red,153;green,92;blue,214}, dashed, no head, from=7-3, to=7-7]
	\arrow["{{\:Octal}}"', from=4-4, to=4-6]
	\arrow[color={rgb,255:red,214;green,153;blue,92}, shorten <=3pt, shorten >=3pt, Rightarrow, 2tail reversed, from=6-5, to=4-5]
	\arrow[color={rgb,255:red,214;green,153;blue,92}, shorten <=3pt, shorten >=3pt, Rightarrow, 2tail reversed, from=4-5, to=2-5]
\end{tikzcd}
        \caption{Refinement in the {\bf AbsProc} as natural transformation between abstract processes and natural transformation between different physical realization of addition (The example for decimal to binary to assembly was adapted after \cite{Horsman2014}).}
        \label{fig:Refine}
\end{figure}
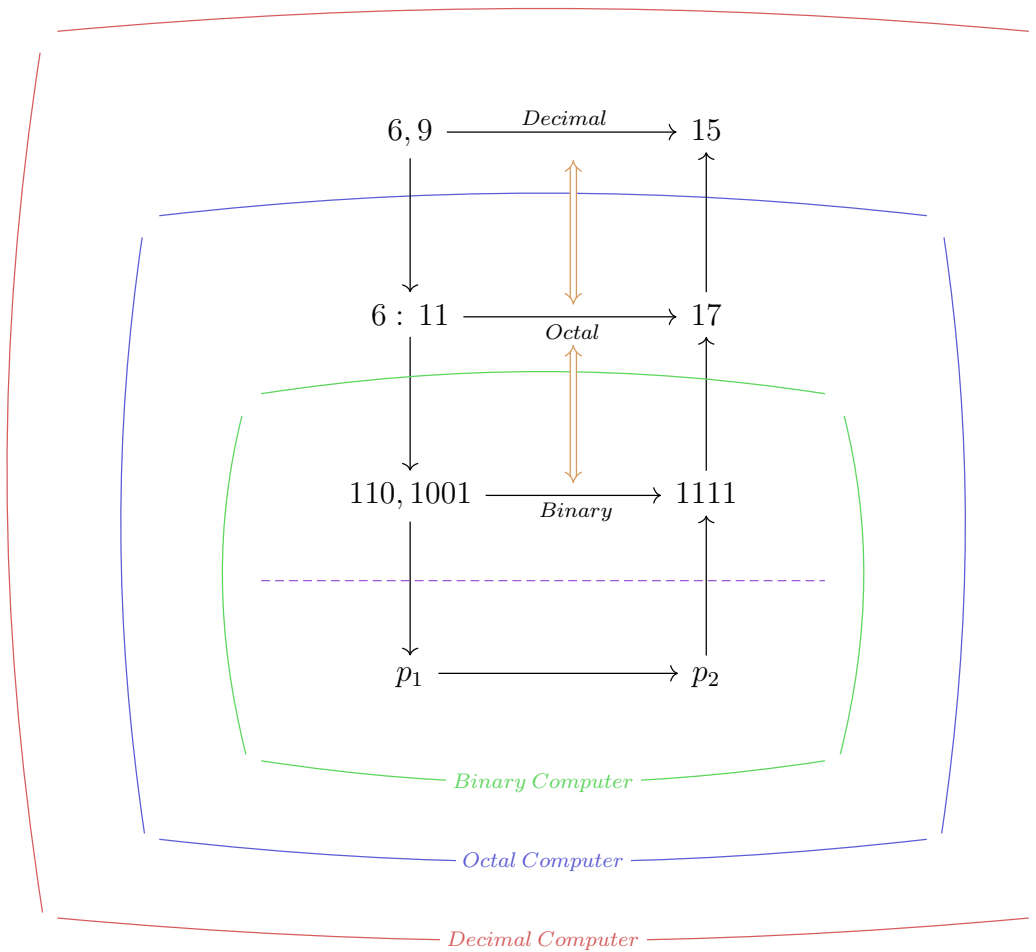

Take the relationship between decimal, octal, and binary adders as an example. This relationship can be redefined through natural transformations, as shown in Figure \ref{fig:Refine}. Here, the physical device could be anything that implements addition, such as voltage fluctuations in transistors (represented by black arrows). Other physical devices, like vacuum tubes, could also serve as equivalent computing machines, directly mapping physical state transitions to the binary abstraction. In all these cases, the intended decimal additions are performed on binary devices via the natural transformation from decimal to binary.

It's also possible to establish a direct functorial relationship between {\bf PhysProc} and {\bf AbsProc}, given that natural transformations form their own category, aligning with the compositional nature of these operations. For instance, a decimal computer could directly use decimal input/output for calculations, eliminating the need for a decimal-to-binary natural transformation. A system like Pascal's mechanical calculator, with 10-tooth input/output wheels and interconnected cogwheels for carry mechanisms, could serve as such a decimal computer. By simply changing the number of teeth on the cogwheel, this addition machine could easily be converted to an octal one, establishing a natural transformation between decimal (Fig.\ref{fig:Refine}:red), octal (Fig.\ref{fig:Refine}:blue), and binary (Fig.\ref{fig:Refine}:green) computers.

\subsubsection{Multiple Realizability Through Adjoint Pairs:}
Our categorical framework extends beyond the basic notion of functorial relationships between {\bf PhysProc} and {\bf AbsProc}. By incorporating the concept of adjoint pairs, it captures equivalences in categories rather than exact identities. This allows for functional isomorphisms between diverse physical systems implementing the same abstraction. 

Functional isomorphism, often described in non-mathematical terms, is a frequently invoked in the philosophy of mind and computational neuroscience. Hillary Putnam proposed that a single mental state or property could correspond to various physical states or properties \cite{Putman1964}, a concept later termed as ``multiple realizability'' \cite{Bickle2020}. David Marr's Tri-Level Hypothesis \footnote{David Marr's Tri-Level Hypothesis consists of: the computational level (the function that the system performs), the algorithmic level (the procedure or representation of the function), and the implementational level (how the computation or algorithm is physically realized).} 
suggests that a single computation can be physically implemented in multiple ways \cite{Marr1976,Marr2010,Vaina2016}. 

\begin{figure}[H]\scalebox{.4}
\centering
\[\begin{tikzcd}
	&&&&& {{\mathcal{R_T}(p'')}} \\
	{a} &&&& {{\mathcal{R_T}(p')}} \\
	&&& {{\mathcal{R_T}(p)}} \\
	{{\widetilde{\mathcal{R}}_T}(a)} && p \\
	&& {p'} \\
	&& {p''}
	\arrow["{{\widetilde{\mathcal{R}}_T}}"{description}, curve={height=12pt}, from=2-1, to=4-1]
	\arrow["f"{description}, color={rgb,255:red,92;green,92;blue,214}, curve={height=6pt}, squiggly, from=4-1, to=4-3]
	\arrow["{f'}"{description}, color={rgb,255:red,92;green,92;blue,214}, squiggly, from=4-1, to=5-3]
	\arrow["{f''}"{description}, color={rgb,255:red,92;green,92;blue,214}, curve={height=-6pt}, squiggly, from=4-1, to=6-3]
	\arrow["{{\mathcal{R_T}(f)}}"{description}, color={rgb,255:red,214;green,92;blue,92}, curve={height=-6pt}, squiggly, from=2-1, to=3-4]
	\arrow["{{\mathcal{R_T}}}"{description}, curve={height=6pt}, from=4-3, to=3-4]
	\arrow["{{\mathcal{R_T}}}"{description}, curve={height=18pt}, from=5-3, to=2-5]
	\arrow["{{\mathcal{R_T}}}"{description}, curve={height=30pt}, from=6-3, to=1-6]
	\arrow["{{\mathcal{R_T}(f'')}}"{description}, color={rgb,255:red,214;green,92;blue,92}, curve={height=6pt}, squiggly, from=2-1, to=1-6]
	\arrow["{{\mathcal{R_T}(f')}}"{description}, color={rgb,255:red,214;green,92;blue,92}, squiggly, from=2-1, to=2-5]
\end{tikzcd}\]
        \caption{Adjoint pair shows multiple realizability of a given abstraction in different physical systems.}
        \label{fig:Multiple}
\end{figure}
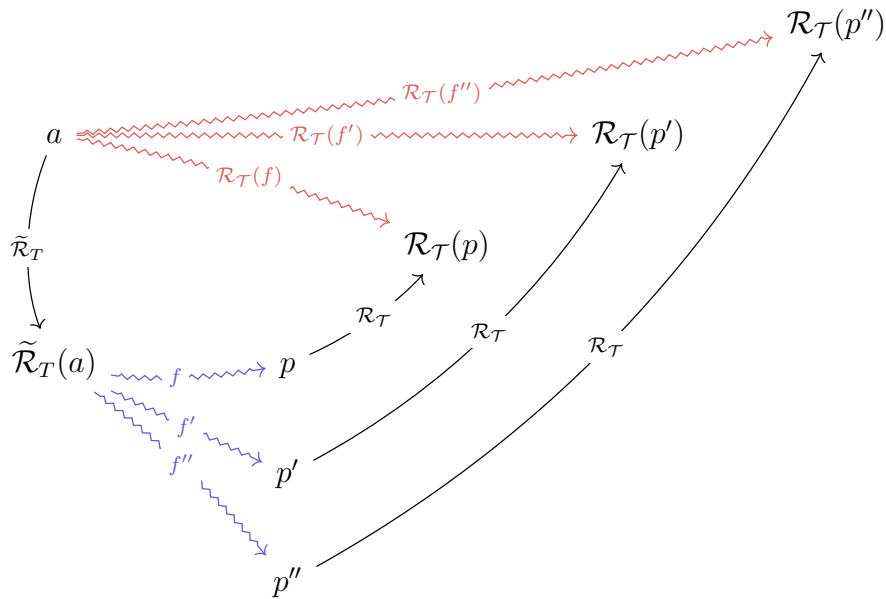

Categorical adjunction, as depicted by the left and right adjoints (mapping between {\bf AbsProc} and {\bf PhysProc}) in Fig. \ref{fig:Multiple}, offers a mathematical interpretation of functional isomorphism in the study of physical computation, demonstrating how different physical systems can realize the same abstract computation. This mathematical framework can serve as a potent tool in describing systems like the brain.

\subsubsection{Nested Composition:}
Exploring the categorical framework further, we introduce the concept of \emph{Nested} composition, which allows for complex, layered interactions between physical and abstract processes. This is exemplified through the nested composition of {\bf PhysProc} to {\bf AbsProc}, demonstrating how multiple layers of abstraction can be sequentially applied to physical systems. 

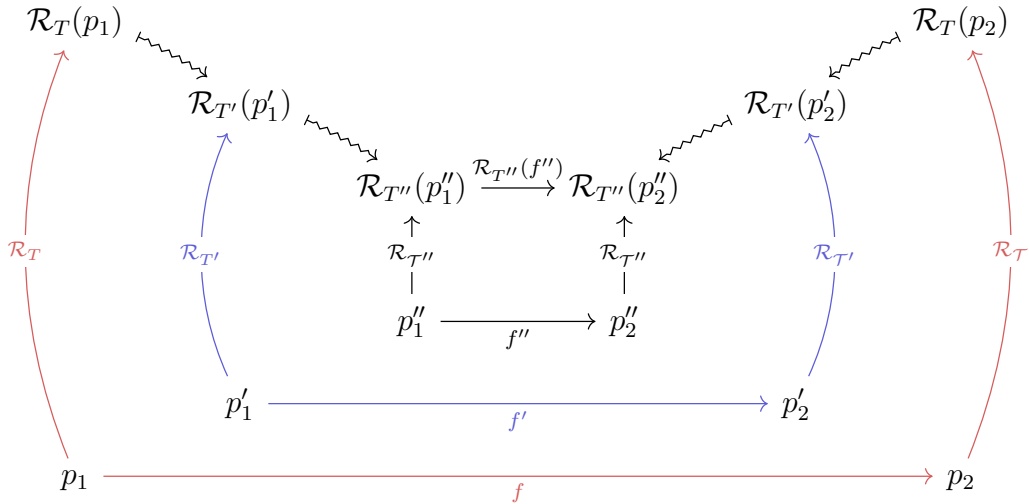
\begin{figure}[H]
\scalebox{.5}
     \centering
\[\begin{tikzcd}[column sep=small, row sep=small]  
	& {{\mathcal R_T(p_1)}} &&&&&& {{\mathcal R_T(p_2)}} \\
	&& {{\mathcal R_{T'}(p'_1)}} &&&& {{\mathcal R_{T'}(p'_2)}} \\
	&&& {{\mathcal R_{T''}(p''_1)}} && {{\mathcal R_{T''}(p''_2)}} \\
	{} &&&&&&&& {} \\
	&&& {p''_1} && {p''_2} \\
	&& {p'_1} &&&& {p'_2 } \\
	& {p_1} &&&&&& {p_2}
	\arrow["{{\mathcal R_{T''}(f'')}}", from=3-4, to=3-6]
	\arrow["{f''}"', from=5-4, to=5-6]
	\arrow["{{\mathcal{R_{T''}}}}"{description}, from=5-4, to=3-4]
	\arrow["{{\mathcal{R_{T''}}}}"{description}, from=5-6, to=3-6]
	\arrow["{{\mathcal R_{T'}}}"{description}, color={rgb,255:red,92;green,92;blue,214}, curve={height=-18pt}, from=6-3, to=2-3]
	\arrow["{f'}"', color={rgb,255:red,92;green,92;blue,214}, from=6-3, to=6-7]
	\arrow["{{\mathcal{R_{T'}}}}"{description}, color={rgb,255:red,92;green,92;blue,214}, curve={height=18pt}, from=6-7, to=2-7]
	\arrow["f"', color={rgb,255:red,214;green,92;blue,92}, from=7-2, to=7-8]
	\arrow["{{\mathcal{R_T}}}"{description}, color={rgb,255:red,214;green,92;blue,92}, curve={height=24pt}, from=7-8, to=1-8]
	\arrow[curve={height=-6pt}, squiggly, maps to, from=1-2, to=2-3]
	\arrow[curve={height=6pt}, squiggly, maps to, from=1-8, to=2-7]
	\arrow[curve={height=-6pt}, squiggly, maps to, from=2-3, to=3-4]
	\arrow[curve={height=6pt}, squiggly, maps to, from=2-7, to=3-6]
	\arrow["{{\mathcal R_T}}"{description}, color={rgb,255:red,214;green,92;blue,92}, curve={height=-24pt}, from=7-2, to=1-2]
\end{tikzcd}\]
        \caption{Nested computation as represented by natural transformations (covariant case).}
        \label{fig:NestedCov}
\end{figure}

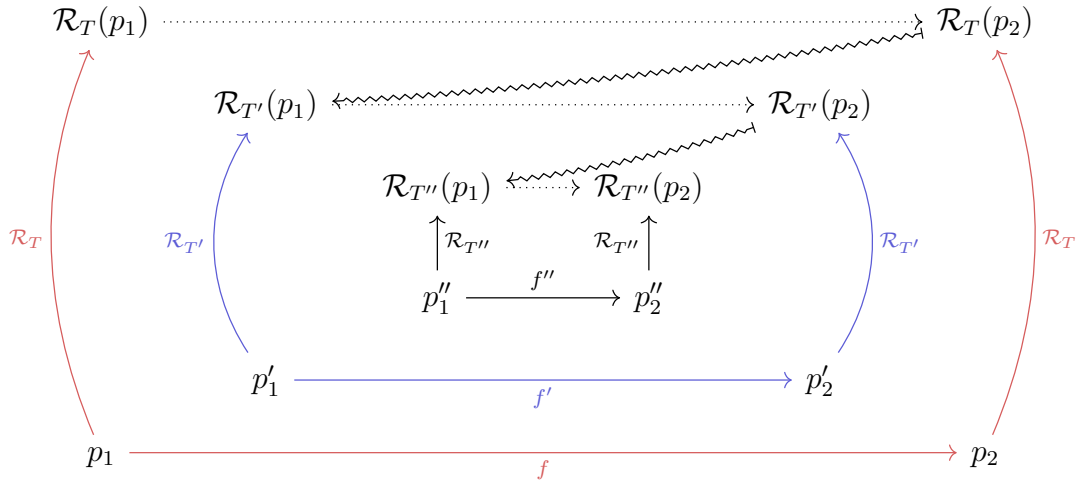
\begin{figure}[H]
\scalebox{.5}
     \centering
\[\begin{tikzcd}[column sep=small, row sep=small] 
	{{\mathcal R_T(p_1)}} &&&&&& {{\mathcal R_T(p_2)}} \\
	& {{\mathcal R_{T'}(p_1)}} &&&& {{\mathcal R_{T'}(p_2)}} \\
	&& {{\mathcal R_{T''}(p_1)}} && {{\mathcal R_{T''}(p_2)}} \\
	\\
	&& {p''_1} && {p''_2} \\
	& {p'_1} &&&& {p'_2} \\
	{p_1} &&&&&& {p_2}
	\arrow["f"', color={rgb,255:red,214;green,92;blue,92}, from=7-1, to=7-7]
	\arrow["{{\mathcal R_T}}", color={rgb,255:red,214;green,92;blue,92}, curve={height=-24pt}, from=7-1, to=1-1]
	\arrow["{{\mathcal R_T}}"', color={rgb,255:red,214;green,92;blue,92}, curve={height=24pt}, from=7-7, to=1-7]
	\arrow[dotted, from=1-1, to=1-7]
	\arrow[dotted, from=2-2, to=2-6]
	\arrow["{{\mathcal R_{T'}}}", color={rgb,255:red,92;green,92;blue,214}, curve={height=-24pt}, from=6-2, to=2-2]
	\arrow["{f'}"', color={rgb,255:red,92;green,92;blue,214}, from=6-2, to=6-6]
	\arrow["{{\mathcal R_{T'}}}"', color={rgb,255:red,92;green,92;blue,214}, curve={height=24pt}, from=6-6, to=2-6]
	\arrow[dotted, from=3-3, to=3-5]
	\arrow["{{\mathcal R_{T''}}}"', from=5-3, to=3-3]
	\arrow["{f''}", from=5-3, to=5-5]
	\arrow["{{\mathcal R_{T''}}}", from=5-5, to=3-5]
	\arrow[curve={height=-6pt}, squiggly, maps to, from=1-7, to=2-2]
	\arrow[curve={height=-6pt}, squiggly, maps to, from=2-6, to=3-3]
\end{tikzcd}\] 
        \caption{Nested computation as represented by natural transformations (contravariant case).}
        \label{fig:NestedContra}
\end{figure}

Figures \ref{fig:NestedCov} and \ref{fig:NestedContra} illustrate an example of nested composition. This involves a functorial mapping of {\bf PhysProc} by a physical system $P$ with state transitions $p_1$ $\rightarrow$ $p_2$, which is then mapped to {\bf AbsProc}. This {\bf AbsProc} is then taken by another physical system $P'$ (with state transitions $p'_1$ $\rightarrow$ $p'_2$) to form a new {\bf AbsProc}. This new {\bf AbsProc} is then functorially mapped to a third system $P''$ (with state transitions $p''_1$ $\rightarrow$ $p''_2$).

This composition demonstrates how a series of abstractions, implemented by interacting physical systems, can execute complex functions. For instance, consider a physical system $P$ with $N$ degrees of freedom connected to another physical system $P'$ with $N'$ degrees of freedom, where $N>N'$. If this second system is then connected to a third system $P''$ with even fewer degrees of freedom ($N'>N''$), the images below use natural transformations to depict the relationship between lower-dimensional embedding and higher levels of abstraction in this nested compositional structure.

In essence, linked open dynamical systems with joint variables can be viewed as nested compositional structures.

 

Our categorical approach can be extended to modular assemblies, similar to biological systems, where both hierarchical and nested compositions are combined. This framework provides a robust tool for the formal analysis of complex systems with many scales of interactions. Biological systems serve as prime examples of this form of modular assembly. Thus, the category theory framework we've presented here emerges as a powerful tool in the formal examination of intricate cellular and subcellular processes.

\section{Discussion: Implications of a categorical perspective on computation}
\subsection{Representation and computation}

The mechanistic/robust mapping account \cite{Piccinini2010, Piccinini2021} and AR theory \cite{Horsman2014,Horsman2018} diverge significantly in their approaches to computation and representation. The mechanistic account, rooted in a mechanical philosophy of mind, advocates a non-representational view of computation, focusing on purposeful (teleological) organization of physical components. Conversely, AR theory posits that computation necessitates representation, yet faces criticism for its vague mapping concept. 

Interestingly, both frameworks converge on the medium independence of computation, acknowledging that a given computation can be physically implemented in multiple ways. This shared view emerges despite their different foundational approaches, partly because neither adhere to the requisite limits of their chosen framework.
The mechanistic account progresses from a {\em specific computation} $\rightarrow$ {\em operations} $\rightarrow$ {\em natural system (specifically brain)} and yet tries to extend this to the decision problem on physical computation. AR attempts to follow {\em Natural systems} $\rightarrow$ {\em operations} $\rightarrow$ {\em mathematical abstraction} and yet resorts to a representation that does not impose strict physical-computational equivalence. 

Our functorial perspective resolves this issue by providing a structured mapping from {\bf PhysProc} to {\bf AbsProc}. This approach avoids the arbitrary nature of representation in AR and bypasses pre-loaded notions of mind inherent in semantic or syntactic accounts, offering a clear pathway to define physical-computational equivalence.

\subsection{Computation and scale}
In our categorical framework, the notion of medium independence and multiple realizability is re-examined, moving away from the traditional trajectory of defining computational nature in physical systems. Here, the organization of matter and degrees of freedom in a physical system dictate the types of possible computations, distinguishing between the implementation of computable functions and the inherent computational capabilities of a system.

To address ``how many different ways a given computable function can be physically implemented," one must recognize the required operations for the function and then identify or construct a physical system with the necessary degrees of freedom to achieve physical-computational equivalence. Conversely, to answer ``when does a physical system compute," the focus shifts to identifying the system's degrees of freedom and determining which computable functions can be executed using these constraints.

The functorial perspective we propose provides a meaningful constraint for physical-computational equivalence, highlighting that the assessment of physical computation should not rely on arbitrary definitions of representation or pre-loaded notions of mind. This perspective reveals the scale-dependency of physical computation, where the morphisms and objects in {\bf PhysProc} and their functorial mapping to {\bf AbsProc} vary depending on the scale of observation.

An illustrative example is the spin qubit computer \footnote{Spin qubit computer (also known as Loss--DiVicenzo quantum computer) was first theoretically suggested in 1998 \cite{Loss1998}. The first experimental demonstration of coherent control of a single-atom electron spin qubit in silicon \cite{Pla2012} has led to the possibility of building a scalable quantum computer. For a review of recent advances and the path to scalable quantum computing with silicon spin qubits, see \cite{Vinet2021}. In a simple form of spin qubit computer, a double quantum dot (a semiconductor particle a few nanometres in size) with two electrons can have L (left) and R (right) spin. A narrow junction permits swapping operation between the two electrons. The electrons confined in this quantum dots system have an intrinsic spin $\frac{1}{2}$ reflecting the number of symmetrical facets in one full rotation. Here the fundamental property of the quantum dot electrons, i.e. spin $\frac{1}{2}$, is the degrees of freedom of the system where information can be registered, manipulated and read. These are equivalent to write (encode), operation and read (decode) as morphisms applied to objects (electron). It has been shown that such quantum dots can be used to build simple quantum logic gates (controlled-NOT) \cite{Bareno1995}.}, where the fundamental property of quantum dot electrons, i.e., spin $\frac{1}{2}$, represents the degrees of freedom for information processing. These quantum properties form the basis for constructing quantum logic gates and conditional quantum dynamics, serving as building blocks for quantum computing systems.

Quantum Dot Cellular Automata (QCA) exemplify a blend of discrete dynamical systems and spin qubit computing principles \footnote{QCA was first theoretically proposed in 1993 \cite{Lent1993} as an edge driven computing system. In QCA, only the edge of the CA array acts as the interface for I/O (write/read) to the system. The first experimental version of this transistorless computing device was a six-dot quantum-dot cellular system (four-dot QCA cell and two electrometer dots) built in 1998 \cite{Snider1998}. In QCAs, just like spin qubit computers, the individual quantum dots have the intrinsic spin $\frac{1}{2}$ and the associated degrees of freedom. The design of the QCA system is as such that boundary conditions (I/O) and the interactions between the internal nodes of the array create the state configurations that produce the computational result. QCAs can be used to construct and interconnect AND gates, OR gates, and inverters}. In QCAs, the intrinsic spin of quantum dots and the design of the system enable the creation of state configurations necessary for computation, illustrating how computational capabilities can manifest at different scales - from macro-scale discrete dynamical systems to micro-scale quantum spins.

Through our categorical framework, the permissible distinguishable states of a physical system at various scales are examined, recognizing that constraints and interscale interactions define these states. The functorial relationship between {\bf PhysProc} and {\bf AbsProc} underpins the physical-computational equivalence in the computing cycle, offering a nuanced view of how physical states translate into computational processes across different scales.

\subsection{Dynamics, information and causality}
AR theory's approach, lacking causal constraints on the mapping between physical and abstract domains \cite{Fletcher2018}, contrasts with the mechanistic/robust mapping account's teleological perspective \cite{Piccinini2010, Piccinini2021}. This divergence in approach stems from AR's resemblance to the simple mapping account and mechanistic account's philosophical roots in framing physical computation.

Causal and counterfactual additions to mapping account variants were introduced to constrain interpretations leading to pancomputationalism. The counterfactual account requires an isomorphism between physical states' counterfactual relations and computational states' counterfactual relations \cite{Campbell2021, Copeland1996, Chalmers1996}. Meanwhile, the causal account mandates an isomorphism between physical states' causal relations and computational states' counterfactual relations \cite{Chrisley1994, Chalmers1996, Scheutz1999, Schnitzer2002}. In both accounts, it is the (counterfactual or causal) relation in the physical system that is required to follow the (counterfactual or causal) relation in the computational domain. This imposing directionality in these variants of mapping account is in contrast with the nature of what we are trying to establish, i.e. examining ``when a physical system computes?''. 

In our categorical framework, these causal and counterfactual relationships are inherently embedded in the \emph{functorial relationship} between {\bf PhysProc} and {\bf AbsProc}. This reflects in two conditions:
\begin{enumerate}
    \item Causal: A state transition in the physical system ($p_1 \rightarrow p_2$) that causes $p_1$ to end in $p_2$, is mirrored by a state transition from $c_1$ (corresponding functorial map of $p_1$) to $c_2$ (functorial map of $p_2$), preserving the causal structure.
    \item Counterfactual: if the computational state is in $c_1$ (that is mapped to from the physical state $p_1$), then the computational state should have gone to $c_2$ (which the evolution of $p_1$, i.e. $p_2$ maps to).
\end{enumerate}
Consequently, the functorial perspective inherently integrates causal and counterfactual elements, ensuring that {\bf AbsProc} are `physically computable' only if the computational operations align with permissible physical state transitions in {\bf PhysProc}. 

These aspects of the categorical perspective of physical computation, combined with representation and scale, point to the foundational relation of the notion of physical computation with pattern formation and information transfer in physical systems. The functorial view that discrete computation {\bf AbsProc} is embedded in continuous dynamical physical systems {\bf PhysProc}, is tightly connected with ``computational mechanics'' and  $\epsilon$-machine description of pattern formation in natural systems \cite{Shalizi2001,Crutchfield2012}. Note that this notion of physical information is rather different from the semantic-latent notion of information \cite{Isaac2019} that considers the mutual information to add a layer to the content-void description of uncertainty in the original formulation of entropic information \cite{Shannon1948}. 

In the functorial perspective of computation, the morphism (dynamics) of natural objects are the rules of continuous dynamical systems. Information, here, is rooted in the state transitions such that two discernible states of the {\bf PhysProc}, $p_1$ and $p_2$ carry descriptions of the generative mechanism of patterns in the physical system. Mapping this physical evolution that carries the causal and counterfactual properties of the physical system, transforms this physical information to the representations according to rules coordinated by their functorial relation. Physical computation is the whole process that carries over the causal states and the transition dynamic to the abstract domain. Under this view, the generative mechanism of pattern formation, symmetry breaking and computation are tightly related.

\subsection{Pandora's box of pancomputationalism}
In addressing pancomputationalism, this paper diverges from specific views on the nature of matter and computation \cite{Fredkin1990, Toffoli1982} or matter and information \cite{Wheeler1982}.
Whether at sufficient small scale one can consider all the fundamental variables and state transitions to be of discrete nature -- lending themselves to be captured by cellular automata \cite{Fredkin1990, Toffoli1982} -- or whether the notion of ``it from bit'' frames that the information ontology preceding physical systems \cite{Wheeler1982} have metaphysical flavors. 
The focus here is not on the metaphysical implications of these theories but on the practical aspects of physical computation.

Pancomputationalism, in its loosest form, suggests that every physical system performs every type of computation, provided a mapping to the abstract domain is possible. A more constrained view posits that every physical system performs at least one computation. This perspective, originating from Putnam's simple mapping account, leads to pancomputationalism being either uninteresting or trivial \cite{Piccinini2021}.

To counter unlimited pancomputationalism, various forms of the mapping account have introduced constraints through semantic, causal, or counterfactual restrictions \cite{Fletcher2018}. For example, the causal mapping account \cite{Chrisley1994, Scheutz1999, Chalmers1996, Chalmers1994} suggests that physical systems are capable of performing computations dictated by their causal structure. Similarly, the semantic mapping account \cite{Shagrir2022, Shagrir2006} limits computation to the manipulation of information-bearing representations.The mechanistic/robust mapping accounts \cite{Anderson2023, Piccinini2021} insist on physical-computational \emph{equivalence} beyond simple \emph{mapping}. The robust mapping's ``equivalence'' enforces a restriction to ensure that the computational state is faithfully mirrored in the physical state. The issue with these attempts for resolving pancomputationalism is that they are focused on ``cognitive computing'' rather than ``when does a physical system compute?''. In essence, they are constructed to deal with {\em specific computation} $\rightarrow$ {\em operations} $\rightarrow$ {\em natural system (brain intended)} instead of {\em Natural systems} $\rightarrow$ {\em operations} $\rightarrow$ {\em mathematical abstraction}.

AR theory doesn't provide a reliable defense against pancomputationalism. The somewhat vague notion of mapping at the core of AR theory is similar, at least structurally, to the simple mapping account and it does not inherently invoke any causal or counterfactual relations. Therefore, AR only provides a physical-computational equivalence that does not provide any constraints beyond mapping. AR attempts to avoid pancomputationalism by externalizing the encoding and decoding (to and from the abstract domain) to an external agent; a notion that is further extended to ``agential AR'' \cite{Fletcher2018} to overcome such deficiencies. Albeit, a key issue with AR remains the lack of constraints in the physical-computational equivalence.

Our categorical framework, however, builds upon AR's strengths to provide an objective criterion for examining physical computation. It incorporates both causal and counterfactual constraints in the mapping from {\bf PhysProc} to {\bf AbsProc} and steers clear of cognitive-bound interpretations often found in philosophy of mind-driven accounts of concrete computation.

In our functorial approach, the physical-computational mapping is either relaxed or strictly equivalent, depending on the context. This compositional nature, combined with the ability to scale the computation within the framework itslef, eliminates the need for an external cognitive agent to oversee the dynamics. It posits that a physical system undergoing state transitions becomes a computational device only when these transitions map meaningfully to the abstract process category, aligning with the operations of a computable function.

Additionally, the framework highlights scale and nested composition in physical computation. For instance, an electron spin in isolation does not constitute a computing device. However, when linked to higher-scale read-out and write-in operations, such as in a quantum dot system, it becomes part of a potential computing device  (like quantum dot). Furthermore, if an abstraction of one physical process maps to state transitions of another physical process, the nested composition of these systems can form a computational device. This form of constrained computational equivalence of physical processes provides an ontological scale and composition that captures the type of computation that can be performed by the system.

\subsection{Categorical framework for physical computation}
Turing and Church's approach to computability, as established in response to Hilbert's Entscheidungsproblem \cite{Turing1937, Church1936, Hilbert1902} invokes two basic questions, \emph{What does constitute a computable function?} and \emph{What functions are noncomputable?}, and results in two basic ramifications, \emph{Computable functions constitute only a small fraction of all functions.} and \emph{We don't need to build different computers for different computable functions.} \cite{Copeland2013}. 

While this abstract notion of computation and the physical devices designed to execute such computations (such as digital computers), are clear, the reverse process {\em Natural systems} $\rightarrow$ {\em operations} $\rightarrow$ {\em mathematical abstraction} often leads to complex questions with unclear answers. For instance, does a rock compute its trajectory? \cite{Chalmers1996}, Is slime mould a computer? \cite{Adamatzky2010}, can a chemical wave be considered a computing system?\cite{Kuhnert1989} Or do bacteria compute?\cite{Benjacob2009}. Here, our focus shifts from the conventional computational models to exploring the computational nature of physical systems, like bacteria, slime moulds, and chemical waves \cite{Chalmers1996, Adamatzky2010, Kuhnert1989, Benjacob2009}.

It’s important to clarify that these questions are not related to designing physical computing systems other than digital computers. Some entities in what is termed as ``unconventional computing” \cite{Teuscher2014} are inspired by nature to create energy-efficient hardware, such as neuromorphic or neuro-inspired designs \cite{Markovic2020,Darwish2018}, or to use biological materials for computations, like DNA computing \cite{Adelman1994}. These attempts aim to use innovative architectural designs and new materials to surpass the limitations of conventional silicon-based computing. However, the trajectory in these domains remains from the computable function to computers.
By employing category theory, we've extended AR theory's conceptual framework \cite{Horsman2014, Horsman2018} to address the critical questions about the nature of computation in physical systems. This approach offers new insights into \emph{What physical system can be considered a computer?} and \emph{Under what constraints, does a physical system compute?}

Key aspects of our categorical framework include:
\begin{enumerate}
    \item Formalization is achieved through categorical \emph{commutative diagrams}, where objects and morphisms represent physical and abstract systems and the operations within and between them. 
    \item \emph{Compositionality} of commuting diagrams provides the tools to probe complex computations at multiple scales.
    \item Adopting a \emph{functorial perspective}, the physical-computational mapping takes a formal definition. The abstract representation can naturally be seen as a functor between the categories phsyical systems {\bf PhysProc} and abstract representations {\bf AbsProc}.
    \item \emph{Natural transformations} guides us to see multiple refinement of {\bf AbsProc}.
    \item \emph{Adjunctions}: we can relax the requirement that the $\mathcal{{R_T}}$ has an inverse $\widetilde{\mathcal{R_T}}$ and consider the case when $(\mathcal{{R_T}},\widetilde{\mathcal{R_T}})$  is an adjoint pair. The intepretation is that  multiple realizability of the same foundational computational process can be instantiated in different physical machines {\bf PhysProc}.   
\end{enumerate}

Overall, this categorical perspective enriches our understanding of physical computation, distinguishing it from mere technological advancements and redefining our concept of computational systems.

The categorical formalism of physical computation introduces several significant implications. It moves beyond subjective definitions of computing systems, offering an objective framework to determine when a physical system computes. Hierarchical and nested compositions in this framework enable the exploration of multiple realizability, showing how various physical devices can implement the same computational processes through different levels of abstraction. This approach also aids in understanding complex, modular systems like living organisms. Additionally, it underscores the importance of scale and symmetry breaking in the context of information transfer, highlighting the dispositional properties of physical systems across different scales. 

\subsection{Precise definitions and robust explanations}
Our categorical approach to physical computation significantly enhances the traditional views of computational systems. This framework is not subject to whimsical reinterpretation, thereby providing a stable and robust explanation of physical computation phenomena. It goes beyond the oversimplification seen in some computational theories regarding physical-computational equivalence. Through strict functorial mappings and the concept of adjoint pairs, we can rigorously define what constitutes a computing device, as opposed to a mere physical system going through change in its microstate configuration.

From an epistemological perspective, the strength of a good paradigm and scientific explanation lies not only in the accuracy of their descriptions and approximations but also in the pitfalls they avoid. This avoidance is indicative of robust and hard-to-vary explanations. If a theory can be modified to accommodate any observation, it is a poor explanation of the natural phenomenon. For instance, consider computational modeling, which approximates the dynamical evolution of physical systems. The assumption of physical-computational equivalence, which is extendable in some forms of concrete computation mapping, leads to the interpretation that the physical system emulates the abstract notion of computation. However, a sound theory should not allow the model of the system to be considered as a computing device performing the computation. The categorical framework of physical computation avoids this pitfall by providing precise definitions of categories, their objects, morphisms, functorial maps between abstract and physical processes, and the adjoint of physical-to-abstract and abstract-to-physical mapping. 

Consider a propelled object. It will traverse its trajectory regardless of the presence of a simulator or observer, or the quality of their theories/models. The propelled object cannot be a device computing all these non-existent, poor, or good theories. The categorical frame provides a clear definition of the physical system and the state transitions it undergoes. It requires a functorial mapping and does not fall for the physical-computational equivalence that does not meet the requirements for the category of abstract processes and the natural transformation between physical and abstract domains. 

In contrast, when a set of balls (as in a conservative logic billiard ball computer) is constrained to perform logical operations as a reversible mechanical computer, the categorical frame can provide an accurate description of physical computing based on Newtonian dynamics. Despite the hypothetical nature of such a physical computer due to the requirement of frictionless mechanics, the mapping relation between the physical process and abstract process and the write-in and read-out of the system have a clear description in the functorial commutative diagram of physical computation. Simulations of the Billiard-ball computer through other Reversible Cellular Automata \cite{Margolus1984,Durand-Lose2002} also align well with the proposed categorical frame. In this case, the equivalence of computing of the physical Billiard-ball computer and the discrete reversible cellular automata can be reasonably conceived as a computational refinement by the means of natural transformation. The proven Turing universality of both the Billiard-ball model and reversible automata simulating the billiard ball model \cite{Durand-Lose2002} further supports this argument.

In summary, our categorical framework provides a concrete and lucid distinction between physical systems and computational devices, ensuring that physical-computational equivalence is not misinterpreted or overly simplified. This robust approach underlines the importance of strict mappings and transformations, setting a clear boundary between genuine computational processes and simple physical occurrences.

\section{Conclusion}
This paper tackles the complex issue of identifying physical systems as computers through a category theory framework. This approach offers a formal framework that naturally encapsulates the process of physical computation. This categorical framework offers a concrete and well-structured definition of physical-to-abstract mapping. It employs composition, adjunction, and natural transformations to rigorously model refinements, multiple realizability, and computation at scale.

Given the high level of abstraction in our representation model, it’s plausible to extend this approach to biological systems. This would prevent misguided comparisons of brains or living systems to computers. Instead, our framework allows us to examine these systems and determine whether they possess, in whole or in part, the characteristics of physical computing systems. We plan to explore this in future work.

Furthermore, the Church--Turing-Deutsch principle, aligning with our framework, posits computation as inherently physical \cite{Deutsch1985}. It suggests universal computing devices could simulate all physical processes, a notion originally influenced by Gandy, Turing's student. This principle invites further exploration of the theoretical limits of ideal physical computing machines, particularly their relation to category theory. Interestingly, Gandy's machines have attracted a category-theoretic axiomatic treatment of update rules to finite objects \cite{Razavi2019}. Future work will deepen our understanding of these relationships and the role of our categorical model in physical computation.

\ack{N.D would like to express gratitude to the organizers and attendees of the \textit{``What is biological computation?''} workshop at \textit{Santa Fe Institute}.  Some of the ideas presented in this work were inspired by discussions that took place during the workshop. The authors wish to thank David Spivak and Gualtiero Piccinini for helpful comments on the manuscript.}

\subsubsection*{Funding and Competing Interest statement:}
The authors did not receive any specific funding for this work and declare no financial or competing interest.

\section*{Appendix: A brief note on Category Theory}
A \textit{category} is composed of a set of \textit{objects} and \textit{morphisms} (also known as arrows) that connect these objects. Given two objects, $a$ and $b$, a morphism $f$ between them can be represented as either $f: a \rightarrow b$ or $a \stackrel{f}{\rightarrow} b$.

A category adheres to three fundamental axioms:

\begin{enumerate}
\item \textit{Identity}: Each object $a$ has an identity morphism, denoted as $a \stackrel{1_a}{\rightarrow} a$.
\item \textit{Composition}: Two morphisms, $a \stackrel{f}{\rightarrow} b$ and $b \stackrel{g}{\rightarrow} c$, can be composed to form a unique morphism $a \stackrel{g\circ f}{\rightarrow} c$. Here, $\circ$ signifies the operation of composition.
\item \textit{Associativity}: Paths of morphisms compose uniquely. For instance, given three arrows $f: a \rightarrow b$, $g: b \rightarrow c$, and $h: c \rightarrow d$, we have $h \circ (g \circ f) = (h \circ g) \circ f$, resulting in the same morphism from $a$ to $d$.
\end{enumerate}

For a more in-depth exploration of category-theoretical concepts used in this paper, please refer to \cite{Maclane2013, Fong2019}.

A \textit{functor} $F$ is a mapping between two categories, say $\mathbf{A}$ and $\mathbf{B}$. It maps objects to objects and morphisms to morphisms. If a morphism $a \rightarrow a^{\prime}$ in $\mathbf{A}$ is mapped by $F$ to a morphism $F(a) \rightarrow F(a^{\prime})$ in $\mathbf{B}$, then $F$ is termed \textit{covariant}. Conversely, if $F(a \rightarrow a^{\prime})$ maps to $F(a) \leftarrow F(a^{\prime})$, $F$ is termed \textit{contravariant}. A contravariant functor can be viewed as a covariant functor $F: \mathbf{A}^{Op} \rightarrow \mathbf{B}$, where $\mathbf{A}^{Op}$ is derived from $\mathbf{A}$ by reversing all its morphisms.

A particularly significant type of category has \textit{functors} as its objects and \textit{natural transformations} as its arrows. Specifically, given two categories $\mathcal{C}$ and $\mathcal{D}$, the functor category $\mathcal{D}^\mathcal{C}$ is defined as follows:

\begin{itemize}
\item The objects of $\mathcal{D}^\mathcal{C}$ are all functors $\mathcal{C}\rightarrow\mathcal{D}$.
\item The arrows of $\mathcal{D}^\mathcal{C}$ are all natural transformations between functors $\mathcal{C}\rightarrow\mathcal{D}$.
\end{itemize}

Natural transformations serve as morphisms between functors. Suppose we have two functors, $F\in\mathcal{D}^\mathcal{C}$ and $G\in\mathcal{D}^\mathcal{C}$. A natural transformation between $F$ and $G$ is a set of morphisms, denoted as $\eta_O$, which are parameterized by the objects $O\in\mathcal{C}$.

This transformation ensures that for any two objects $A$ and $B$ in $\mathcal{C}$ that are linked by a morphism $f$, the following diagram commutes:

\[
\xymatrix{
F(A)\ar[rr]^{F(f)}\ar[dd]_{\eta_{A}} & & F(B)\ar[dd]^{\eta_B}\\
 & & \\
G(A)\ar[rr]^{G(f)} & &  G(B) 
}
\]

In this diagram, the horizontal arrows represent the action of the functors on the morphism $f$, and the vertical arrows represent the natural transformation $\eta$. The diagram commutes if you can travel from $F(A)$ to $G(B)$ along any path and the result is the same. This is a fundamental property of natural transformations in category theory.



The final concept we discussed in this paper is the \textit{adjoint pair}. This involves two functors, $F$ and $G$, defined as follows:

\begin{itemize}
\item $F$ maps from category $\mathcal{D}$ to category $\mathcal{C}$: 
\[F:\mathcal{D}\longrightarrow \mathcal{C}\]
\item $G$ maps from category $\mathcal{C}$ to category $\mathcal{D}$: 
\[G:\mathcal{C}\longrightarrow \mathcal{D}\]
\end{itemize}

$F$ and $G$ form an adjoint pair if there exist natural transformations:

\begin{itemize}
\item $\eta$, which maps the identity endofunctor on category $\mathcal{C}$ to the composition of functors $F$ and $G$: 
\[\eta: 1_{\mathcal{C}}\longrightarrow F\circ G\]
\item $\epsilon$, which maps the identity endofunctor on category $\mathcal{D}$ to the composition of functors $G$ and $F$: 
\[\epsilon: 1_{\mathcal{D}}\longrightarrow G\circ F\]
\end{itemize}

These transformations must satisfy the following for all objects $c$ in $\mathcal{C}$ and all objects $d$ in $\mathcal{D}$:

\begin{itemize}
\item For $G(c)$: 
\[1_{G(c)}:G(c)\xrightarrow[G(\eta_c)]{} G(F(G((c))))\xrightarrow[\epsilon_{G(c)}]{} G(c)\]
\item For $F(d)$: 
\[1_{F(d)}:F(d)\xrightarrow[\eta_{F(d)}]{} F(G(F((d))))\xrightarrow[F(\epsilon_d)]{} F(d)\]
\end{itemize}

Here, $1_{\mathcal{C}}$ represents the identity endofunctor on category $\mathcal{C}$, while $1_c$ denotes the identity morphism on an object $c$.





\section*{References}
\bibliographystyle{plain}
\bibliography{CatComp.bib}
\end{document}